\documentstyle[preprint,prb,aps,epsf,eqsecnum]{revtex}
\begin{document}
\tightenlines

\title{The universal chiral partition function for exclusion statistics}

\author{Alexander Berkovich
\footnote{e-mail alexb@insti.physics.sunysb.edu}}
\author{Barry~M.~McCoy
\footnote{e-mail mccoy@insti.physics.sunysb.edu}}               
\address{ Institute for Theoretical Physics, State University of New York,
 Stony Brook,  NY 11794-3840}
\date{\today}
\preprint{ITPSB-98-50}

\maketitle

\begin{abstract}
We demonstrate the equality between the universal chiral partition
function, which was first found in the context of conformal field theory
and Rogers-Ramanujan identities, and the exclusion statistics
introduced by Haldane in the study  of
the fractional quantum Hall effect. The phenomena of multiple
representations of the same conformal field theory by different sets
of exclusion statistics is discussed in the context of the ${\hat
u}(1)$ theory of a compactified boson of radius $R.$ 
\end{abstract}
\section{Introduction}

In 1991 Haldane\cite{hald}, in the context of the fractional quantum
Hall effect \cite{lau},
 introduced the following ``{\it statistical 
interaction} $g_{\alpha \beta}$ through the differential relation
\begin{equation}
\Delta d_{\alpha}=-\sum_{\beta}g_{\alpha \beta}\Delta N_{\beta}
\label{haldef}
\end{equation}
where $\{\Delta N_{\beta}\}$ is a set of allowed changes of
the particle numbers at fixed size and boundary conditions'' and
$d_{\alpha}$ is the dimension of the Hilbert space. The key idea
embodied in this definition is that the number of states allowed to a
particle is linearly dependent on the number of particles in the state. 
When $g_{\alpha \beta}=0$ there is no reduction in the number of
states and particles are called bosons whereas when
$g_{\alpha\beta}=\delta_{\alpha \beta}$ the particles obey the Pauli
exclusion principle.
The linear exclusion rule (\ref{haldef}) is ``{\it
considered a generalization of the Pauli principle}'' \cite{hald} and builds on
both previous notions of generalized statistics \cite{lm}-\cite{asw}
used in the fractional quantum Hall effect
and on solutions to integrable models \cite{yy}-\cite{shas}. 
Subsequently this
somewhat general notion was extended and sharpened by Wu \cite{wu}
and others \cite{nw}-\cite{vob} and was reapplied 
to the fractional quantum Hall effect 
by van Elberg and Schoutens \cite{es}. In the course of these
studies \cite{wu}-\cite{es} the linear exclusion relation (\ref{haldef}) 
has come to be referred to as {\it exclusion statistics}.

In 1993, in the context of conformal field theory and the corresponding
integrable lattice models one of the authors and his collaborators \cite{kkmma}-\cite{kkmmb}
introduced (for $y_i=1$) what can be descriptively described as the 

{\bf Universal Chiral Partition Function}
\begin{equation}
S_{\bf B}{{\bf Q}\atopwithdelims[] {\bf A}}({\bf u},{\bf y}|q)=
\sum_{{\bf m=0}\atop {\rm {restrictions} {\bf Q}}}^{\infty}q^{{1\over 2}{\bf
mBm}-{1\over 2}{\bf Am}}\prod_{\alpha=1}^ny_{\alpha}^{m_{\alpha}}
{((1-{\bf B}){\bf m}+{{\bf
u}\over 2})_{\alpha}\atopwithdelims[] m_{\alpha}}
\label{ucpf}
\end{equation}
where for $m$ and $l$ integers the Gaussian polynomials are defined by
\begin{equation}
{l\atopwithdelims[] m}=\cases{{(q)_l\over (q)_m(q)_{l-m}}& for $0\leq
m\leq l$\cr
0& otherwise}~{\rm and}~(a)_l=\prod_{j=0}^{l-1}(1-aq^j)
\end{equation} 
and we note the limiting property
\begin{equation}
\lim _{l\rightarrow \infty}{l\atopwithdelims[] m}={1\over (q)_m}.
\label{binlim}
\end{equation}
Here $\bf m,A,Q$ and $\bf u$ are $n$ component vectors, $\bf B$ is an
$n\times n$ matrix and the restrictions $\bf Q$ on the sum are such
that the arguments of the Gaussian polynomials are integers. 

We call
(\ref{ucpf}) a chiral partition function because it is indeed a grand partition
function for $n$ species of right moving (chiral) 
particles with fugacities $y_{j}$
\begin{equation}
S_{\bf B}{{\bf Q}\atopwithdelims[] {\bf A}}({\bf u},{\bf y}|q)=
\sum_{i}e^{-E_i/k_bT}y_j^{m_j}
\label{part}
\end{equation}
where the sum is over all states $i$ whose energy $E_i$ is given in
terms of single particle energies of particles with a linear
dispersion relation as
\begin{equation}
E_i=\sum_{\alpha=1}^n\sum_{j=1}^{m_{\alpha}}vP_j^{\alpha}.
\end{equation}
Here the single particle momenta are chosen from the set
\begin{equation}
P_j^{\alpha}\in \{P_{\rm min}^{\alpha}({\bf m}),
P_{\rm min}^{\alpha}({\bf m})+{2\pi\over M},\cdots,P_{\rm
max}^{\alpha}({\bf m})\},
\label{count}
\end{equation}
where the Fermi exclusion rule (Pauli principle) holds
\begin{equation}
P_j^{\alpha}\neq P_k^{\alpha}~~{\rm for}~~j\neq k~~{\rm
and~all}~~\alpha,
\label{fermi}
\end{equation}
the minimum and maximum momenta are
\begin{equation}
P_{\rm min}^{\alpha}({\bf m})=
{\pi\over M}[(({\bf B}-1){\bf m})_{\alpha}-A_{\alpha}+1],
~~~~
P_{\rm max}^{\alpha}({\bf m})=-P_{\rm min}^{\alpha}({\bf m})
+{2\pi\over M}({{\bf u}\over 2}-{\bf A})_{\alpha}
\label{mindef}
\end{equation}
and $q=e^{-{2\pi v\over k_bTM}.}$
The expression (\ref{ucpf}) is obtained from (\ref{part}) by use of
the identity
\begin{equation}
\sum_{N=0}^{\infty}Q_m(N;N')q^N=q^{m(m-1)}{N'+1\atopwithdelims[] m}
\end{equation}
where $Q_m(N,N')$ is the number of additive partitions of $N\geq 0$
into $m$ distinct non-negative integers each less than or equal to
$N'.$ We refer to (\ref{count})-(\ref{mindef}) as fermionic counting rules.

We refer to (\ref{ucpf}) as universal because in a long series of
papers (see refs. \cite{lp}-\cite{morebm} and references contained therein) 
it has been seen that the characters of conformal field
theories and branching functions of affine Lie algebras may be
universally written in this form (in the conformal limit
$T\rightarrow 0,~~M\rightarrow\infty$ with $q$ fixed.)

The connection of (\ref{ucpf}) with bosons and fermions is easily seen
by using elementary identities in $q$ series \cite{gr} that date back
to Euler and the q-analogue of the binomial theorem.

For the connection with a free fermion we set $n={\bf B}={\bf A}=1,~{\bf
u}=\infty,$ and consider the unrestricted sum $({\bf Q}=0)$
and use 1.3.16 on page 9 of ref.\cite{gr} to find
\begin{equation}
S_1{0\atopwithdelims[] 1}(\infty,y|q)=\sum_{n=0}^{\infty}{q^{{1\over
2}m(m-1)}y^m\over (q)_m}=(-y)_{\infty}=\prod_{j=0}^{\infty}(1+yq^j).
\end{equation}
The righthand side is manifestly the partition function for a free (chiral)
fermion with a linear dispersion relation. From (\ref{mindef}) we
see that $P_{\rm min}(m)=0$ is independent of $m$ as should be the
case for a free fermion.

For the connection with a free boson we set 
$n=1,~{\bf u}=\infty, {\bf A}=0$ and
consider the unrestricted sum as before, but now we set ${\bf B}=0$ and use
(18) of page xiv of ref.\cite{gr} to find
\begin{equation}
S_0{0\atopwithdelims[] 0}(\infty,y|q)=\sum_{n=0}^{\infty}{y^{m}\over
(q)_m}={1\over (y)_{\infty}}={1\over \prod_{j=0}^{\infty}(1-yq^j)}.
\label{euler}
\end{equation}
The right hand side is manifestly the partition function for a 
free (chiral) boson with a linear dispersion relation. From
(\ref{mindef}) we see that $P_{\rm min}(m)={\pi\over M}(1-m).$ Thus we
see that a particle with a Pauli exclusion principle can indeed have a
bosonic partition function. This cannot not be considered as strange
since the identity (\ref{euler}) has been known for well over 200 years.
The extension to $n$ free bosons with $g_{\alpha \beta}=0$ or 
free fermions with $g_{\alpha \beta}=\delta_{\alpha \beta}$ is obvious.

We can now easily compare the universal chiral partition function
(\ref{ucpf}) with ${\bf u}=\infty$ with the exclusion statistics given by
(\ref{haldef}). The rule (\ref{haldef}) gives a linear exclusion of
states governed by the matrix $g_{\alpha \beta}.$ The universal chiral
partition function (\ref{ucpf}) comes from the state counting formula 
(\ref{count})-(\ref{mindef}) with a linear exclusion of states
governed by the matrix $B_{\alpha \beta}.$ We have just seen that  the
case $g_{\alpha \beta}=B_{\alpha \beta}=0$ gives free bosons and
$g_{\alpha \beta}=B_{\alpha \beta}=\delta_{\alpha \beta}$ gives free
fermions. Therefore the identification is almost obvious that if we set
\begin{equation}
B_{\alpha \beta}=g_{\alpha \beta}
\end{equation}
then the exclusion statistics (\ref{haldef}) of Haldane\cite{hald}
 will lead to the
universal chiral partition function (\ref{ucpf}) with 
${\bf u}=\infty.$ The only difference
in the two formulations is that the rule (\ref{haldef}) 
is excluding states from
a bosonic Fock space while the counting rules
(\ref{count})--(\ref{mindef}) are excluding or adding states to a
fermionic Fock space. The virtue of the fermionic formulation is that
the state counting (\ref{count})--({\ref{mindef}) is very explicit while
for the bosonic construction no such simple explicit formula is known.

\section{The equations of Wu}

The argument just given is very general and is valid for any number of
quasi particles. For the case of one quasi particle, however, a much
more detailed treatment of exclusion statistics was made by
in 1994 by Wu  
\cite{wu} who showed that the energy of systems with 
exclusion statistics in the thermodynamic limit
where $M\rightarrow \infty$ with $T$ fixed is
\begin{equation}
E_{wu}(g)=\rho_{0}\int_0^{\infty}d\epsilon \epsilon n_g(\epsilon)
\label{wuena}
\end{equation}
where
\begin{equation}
n_g(\epsilon)={1\over w(\epsilon)+g}~~{\rm and}~~
w(\epsilon)^g[1+w(\epsilon)]^{1-g}=y^{-1}e^{\epsilon/k_bT}.
\end{equation}
or, equivalently setting $z=\epsilon/k_bT$
\begin{equation}
E_{wu}(g)=\rho_0(k_bT)^2\int_0^{\infty}dz~z~{\bar n}_g(z)
\label{wuen}
\end{equation}
with 
\begin{equation}
{\bar n}_g(z)={1\over {\bar w}(z)+g}~~{\rm and}~~
{\bar w}(z)^g[1+{\bar w}(z)]^{1-g}=y^{-1}e^{z}.
\label{wuhelp}
\end{equation}
In this section we will show the equivalence of these results with
the corresponding results obtained from the universal 
chiral partition function.

For $n=1$ the universal chiral partition function of (\ref{ucpf}) with
${\bf u}\rightarrow \infty,~{\bf A}=0$ and no restrictions ${\bf Q}$
is
\begin{equation}
S_{B}{0\atopwithdelims[] 0}(\infty,y|q)=\sum_{m=0}^{\infty}{q^{{1\over
2}Bm^2}y^m\over (q)_m}.
\label{scalar}
\end{equation}
From the definition of $q$ we see that to study the thermodynamic
limit $M\rightarrow \infty$ with $T$ fixed we need to study the
behavior of (\ref{scalar}) as $q\rightarrow 1.$ This limit has been
studied extensively in ref. \cite{kkmma}-\cite{kkmmb}~ by means of the
method of
steepest descents in the context of the computation of the central charge
of conformal field theory. In the present case we note that 
as $q\rightarrow 1$ the sum in (\ref{scalar}) is dominated by terms
where $m\sim 1/{\rm ln}q^{-1}$ and thus we may write
\begin{eqnarray}
S_B{0\atopwithdelims[] 0}(\infty,y|q)&=&\sum_{m=0}^{\infty}{\rm
exp}\{-{1\over 2}Bm^2{\ln}q^{-1}+m{\rm ln}y-\sum_{j=1}^{m}{\rm
ln}(1-e^{-j{\rm ln}q^{-1}})\}\nonumber \\
&\sim&\sum_{m=0}^{\infty}{\rm exp}\{-{B\over 2}m^2{\rm ln}q^{-1}+
m{\rm ln}y-{1\over {\rm ln}q^{-1}}\int_0^{m{\rm ln}q^{-1}}dt{\rm
ln}(1-e^{-t})\}.
\label{more}
\end{eqnarray}
The sum is dominated by its value at the steepest descents point
determined from
\begin{eqnarray}
0&=&{d\over dm}\{-{B\over 2}m^2{\rm ln}q^{-1}+m{\rm ln}y-{1\over {\rm
ln}q^{-1}}\int_{0}^{m{\rm ln}q^{-1}}dt {\rm ln}(1-e^{-t})\}\nonumber\\
&=&-mB{\rm ln}q^{-1}+{\rm ln}y-{\rm ln}(1-e^{-m{\rm ln}q^{-1}})
\label{steep}
\end{eqnarray}
Thus, setting $x=m{\rm ln}q^{-1}$ we write (\ref{steep}) as
\begin{equation}
ye^{-Bx}=(1-e^{-x})
\label{steepa}
\end{equation}
and using this value of $x$ in (\ref{more}) and recalling the
definition of $q$ we find that as $M\rightarrow \infty$
\begin{equation}
{\rm ln}S_{B}{0\atopwithdelims[] 0}(\infty,y|q)\sim {k_bTM\over 2\pi v} 
\{-{B\over 2}x^2+x{\rm ln}y-\int_0^{x}dt{\rm ln}(1-e^{-t})\}.
\end{equation}
The free energy per site f is defined as
\begin{equation}
f=-k_bT{\rm lim}_{M\rightarrow \infty}{1\over M}{\rm
ln}S_B{0\atopwithdelims[]0}(\infty,y|q)
\end{equation}
and the internal energy per site is
\begin{equation}
E(B)={\partial\over \partial {\beta}}\beta f|_y
\end{equation}
and thus we have
\begin{equation}
E(B)={(k_bT)^2\over 2\pi v}\{-{B\over 2}x^2+x{\rm ln}y-\int_0^{x}dt{\rm
ln}(1-e^{-t})\}.
\end{equation}
This may also be written in terms of the Rogers dilogarithm function
\begin{eqnarray}
L(w)&=&-{1\over 2}\int_0^wdv[{{\rm ln}v\over 1-v}+{{\rm ln}(1-v)\over
v}]\nonumber\\
&=&-\int_0^w dv {{\rm ln}(1-v)\over v}+{1\over 2}{\rm ln}w{\rm ln}(1-w)
=-\int_0^w dv {{\rm ln}v\over 1-v}-{1\over 2}{\rm ln}w{\rm ln}(1-w)
\label{dilog}
\end{eqnarray}
by use of the relation\cite{lewin}
\begin{equation}
L(w)+L(1-w)=L(1)={\pi^2\over 6}
\end{equation}
as
\begin{equation}
E(B)={(k_bT)^2\over 2\pi v}\{{1\over 2}x{\rm ln}y+L(1-e^{-x}\}=
{(k_bT)^2\over 2\pi v}\{{B\over 2}x^2+{1\over 2}x{\rm ln}(1-e^{-x})
+L(1-e^{-x})\}.
\label{usen}
\end{equation}

We will show that if we identify $B$ in (\ref{usen}) with $g$ in
(\ref{wuen}) then
 \begin{equation}
E(g)=E_{wu}(g).
\label{enid}
\end{equation}
We do this by first noting that from (\ref{wuhelp})
\begin{equation}
z=g{\rm ln}{\bar w}+(1-g){\rm ln}(1+{\bar w})+{\rm ln}y~~{\rm
and}~~{d{\bar w}\over dz}={{\bar w}(1-{\bar w})\over {\bar w}+g}.
\label{uwrel}
\end{equation}
Moreover, we see from (\ref{uwrel}) that if $z=\infty$ then  $w=\infty$ 
and by comparing with (\ref{steepa}) (with $B=g$) we see that if $z=0$
then $w=1/(e^x-1).$ Thus we rewrite (\ref{wuen}) using ${\bar w}$ as the
independent variable instead of $z$ and obtain
\begin{eqnarray}
E_{wu}(g)&=&\rho_0(k_bT)^2\int_{1\over e^x-1}^{\infty}{d{\bar w}\over
{\bar w}({\bar w}+1)}\{g{\rm ln}{\bar w}+
(1-g){\rm ln}(1+{\bar w})+{\rm ln}y\}\nonumber\\
&=&\rho_0(k_bT)^2\{gx^2+x{\rm ln}(1-e^{-x})+\int_{1\over
e^x-1}^{\infty}{d{\bar w}\over {\bar w}(1+{\bar w})}
\{g{\rm ln}{\bar w}+(1-g){\rm ln}(1+{\bar w})\}
\end{eqnarray}
where in the last line $y$ has been eliminated in favor of $x$ by use
of (\ref{steepa}) (with $B=g$). In the remaining integral let
\begin{equation}
v={1\over 1+{\bar w}}~~{\rm with}~~d{\bar w}=-{dv\over v^2}
\end{equation}
to find
\begin{eqnarray}
E_{wu}(g)&=&\rho_0(k_bT)^2\{{g\over 2}x^2+x{\rm
ln}(1-e^{-x})-\int_0^{1-e^{-x}}dv{{\rm ln}v\over 1-v}\}\nonumber\\
&=&\rho_0(k_bT)^2[{g\over 2}x^2+{1\over 2}x{\rm
ln}(1-e^{-x})+L(1-e^{-x})]
\end{eqnarray}
where in the last line we have used the definition of $L(w)$ of the
last line of (\ref{dilog}). Upon comparing with (\ref{usen}) we see
that (\ref{enid}) does indeed hold. Thus we have shown that the energy
given by the universal chiral partition function (\ref{ucpf}) with
$n=1$ and $u=\infty$ is identical with the energy computed by Wu\cite{wu} 
for the exclusion statistics of Haldane\cite{hald}.

\section{Multiple representations for ${\hat u}(1)$}

Perhaps the most interesting and least understood property of the
universal chiral partition function is that there
are often identities between forms (\ref{ucpf}) with
different ${\bf B}$ matrices which may even have different dimensions.
Thus, for the minimal model $M(3,4)$ there is an identity between the
universal chiral partition function 
with a one dimensional and an eight dimensional $\bf B$ matrix \cite{kkmmb}.
This multiplicity of representations is often identified with the
various integrable perturbation of the conformal field theory.

The identification of the universal chiral partition 
function (\ref{ucpf}) with the exclusion
statistics (\ref{haldef}) means that this same phenomena of a
multiplicity of representations must occur for exclusion statistics
also. Here we
illustrate this multiplicity phenomena for the ${\hat u}(1)$ affine Lie
algebra.

The conformal field theory of affine Lie algebra 
${\hat u}(1)$ appears in several contexts. In one context \cite{dms} it is the
Gaussian model which describes a boson compactified with a radius $R$
with conformal dimensions
\begin{equation}
h_{n,m}={1\over 2}(n/R+mR/2)^2.
\end{equation}
In the context of the XXZ spin chain 
\begin{equation}
H_{XXZ}=\sum_{j=1}^{L}(\sigma^x_j\sigma^x_{j+1}+\sigma^y_j\sigma^y_{j+1}+
\cos \mu \sigma^x_j\sigma^x_{j+1})
 \end{equation}
the radius $R$ (using the normalization of ref.\cite{dms}) is
\begin{equation}
R=\left({2\over 1-\mu/\pi}\right)^{1/2},~~~0\leq \mu\leq \pi.
\end{equation}
When $R^2=2p'/p$ is rational the extended characters are given for $p'>p$
relatively prime as
\cite{dms}
\begin{equation}
\chi^{(pp')}_l(q,y)={1\over
(q)_{\infty}}\sum_{j=-\infty}^{\infty}q^{pp'(j+{l\over
2pp'})^2}y^j,~~~0\leq l\leq 2pp'
\label{uchar}
\end{equation} 
It was shown in ref.\cite{kmm} that if we write $l=p'Q+Q'$ with
$Q'=0,1,\cdots, p'-1$ and $Q\in Z_{2p}$ then 
 $\chi(q,y)$ may be
written in the form (\ref{ucpf}) with $n=2$
\begin{equation}
B=\left(\begin{array}{cc}
{R^2\over 4}&1-{R^2\over 4}\\
1-{R^2\over 4}&{R^2\over 4}\end{array}\right),
A={Q'\over p}(1,-1),~~m_1-m_2\equiv Q({\rm
mod}2p')
\label{ubmat}
\end{equation}
and $y_1=y_2^{-1}=y^{1\over 2p}.$
In this $2\times 2$  matrix {\bf B} 
the two particles appear in a symmetric fashion.

The ${\hat u}(1)$ conformal field theory also appears in the context of
the Cologero-Sutherland model \cite{col}-\cite{sutb}
\begin{equation}
H_{CS}=\sum{i}{p_i^2\over 2}+({\pi\over
L})^2\sum_{i<j}{\beta^2-\beta\over \sin^2(x_i-x_j)\pi/L}
\end{equation}
where now the radius of compactification \cite{ky},\cite{iso} is $R^2=\beta.$
Here  it has been shown \cite{iso},\cite{hik} that there is 
again a two particle basis, 
(now called anyons and superfermions in \cite{iso} and g-ons in \cite{hik})
but now these two particles do not appear symmetrically.
In this context there is an isomorphism with the fractional quantum
Hall effect \cite{wena},\cite{wenb} where the basis particles are called 
electrons and quasi-particles.
In the Cologero-Sutherland model \cite{hik} (and because of the
isomorphism of models also in the fractional quantum Hall effect \cite{es})
the truncated partition sum for g-ons (quasi-electrons for the filling
fraction $\nu =1/g=r/s$ with $r$ and $s$ relatively prime) 
satisfies the recursion relation for integer $L+{s\over 2}$ 
\begin{equation}
X_L(y,q,r,s)=X_{L-r}(y,q,r,s)+yq^{L\over r}X_{L-s}(y,q,r,s). 
\label{recrel}
\end{equation}
and in \cite{hik} it was shown for $r=1$  
that the
universal chiral partition function (\ref{ucpf}) with 
\begin{equation}
n=1,~~~B={s\over r},~~~{u\over 2}={L\over r}+{s\over 2r}-a
\label{special}
\end{equation}
is an exact solution of (\ref{recrel}) for all  $a.$ 

Here we demonstrate this solution by writing
 the specialization (\ref{special}) of (\ref{ucpf}) as
\begin{equation}
F_{L}(y,q,r,s,a)=\sum_{m=0\atop L+s/2-sm\equiv ra ({\rm
mod}~r)}^{\infty}q^{{1\over 2}{s\over r}m^2+am}y^m{{L\over r}-({s\over
r}-1)m+{s\over 2r}-a\atopwithdelims[] m}
\label{newdef}
\end{equation}
and use the recursion relation for Gaussian polynomials
\begin{equation}
{l\atopwithdelims[] m}={l-1\atopwithdelims[] m}+
q^{l-m}{l-1\atopwithdelims[] m-1}
\end{equation}
to obtain
\begin{eqnarray}F_L(y,q,r,s,a)&=&
\sum_{m=0\atop L+s/2-sm\equiv ra ({\rm mod}~r)}^{\infty}q^{{1\over 2}{s\over
r}m^2+am}y^{m}
({{L\over r}-1-({s\over r}-1)m+{s\over 2r}-a\atopwithdelims[]
m}\nonumber \\
&+&q^{{L\over r}+{s\over 2r}-({s\over r}-1)m-a-m}
{{L\over r}-1-({s\over r}-1)m+{s\over 2r}-a\atopwithdelims[] m-1}).
\end{eqnarray}
The first term is recognized as $F_{L-r}(y,q,r,s,a)$ and in the second
term we set $m-1=j$ and thus we obtain
\begin{eqnarray}&~&F_{L}(y,q,r,s,a)=F_{L-r}(y,q,r,s,a)\nonumber\\
&+&\sum_{j=0 \atop L-s+s/2-sj\equiv ra ({\rm mod}~r)}^{\infty}q^{{1\over
2}{s\over r}(j+1)^2+a(j+1)+{L\over r}+{s\over 2r}-{s\over r}(j+1)-a}
y^{j+1}{{L\over r}-1+{s\over 2r}-({s\over
r}-1)(j+1)-a\atopwithdelims[] j}.
\end{eqnarray}
Finally after simplifying the exponent and comparing with the
definition (\ref{newdef}) the second term is recognized
as $yq^{L/r}F_{L-s}(y,q,r,s,a)$ and thus we obtain the final result
\begin{equation}F_L(y,q,r,s,a)=
F_{L-r}(y,q,r,s,a)+yq^{{L\over r}}F_{L-s}(y,q,r,s,a)
\end{equation}
which is precisely the recursion relation (\ref{recrel}).

This derivation makes clear that the restrictions in the sum of the 
universal chiral partition function arise whenever $B$ is
fractional. These restrictions in essence reduce the partition
function to $1/r$ of the unrestricted partition function and thus will
not effect the exponential behavior of the $M\rightarrow \infty$
thermodynamic limit. Consequently these restrictions have no effect
on the computation of the energy done in the preceding section. 

The partition function for the fractional quantum Hall effect
representation of ${\hat u}(1)$ is obtained from the function
$F_{L}(y,q,r,s,a)$ for the electrons and $F_{L}(y,q,s,r,a)$ for the
quasi particle for finite $L$ by means of the following elegant
polynomial identity for 
$s\geq r$ 
\begin{eqnarray}
&~&\sum_{{j=-\infty \atop j\equiv L ({\rm mod}~s),
\equiv -L ({\rm mod}~r)}}^{\infty}
q^{j^2\over 2rs}y^{-{j\over rs}}{({1\over r}+{1\over
s})L+({1\over r}-
{1\over s})j\atopwithdelims[] {L-j\over s}}\nonumber\\
~~~&=&\sum_{a=1-r}^{s-r}q^{a^2\over 2rs}y^{a\over rs}
F_{L-s/2}(y^{1\over r},q,r,s,{a\over
r})F_{L-r/2}(y^{-{1\over s}},q,s,r, \theta(a>0)-a/s)
\label{uone}
\end{eqnarray}
and $\theta(a>0)=1$ if $a>0$ and zero otherwise. 
We note that there are in general $rs$ different
limits as $L\rightarrow \infty$ depending on the congruences $L\equiv
l ({\rm mod} rs)$ with $l=0,1,\cdots,rs-1$. We also note that
(\ref{uone}) is in general not invariant under $y\rightarrow y^{-1}.$ 
When $r=s=1$ the identity (\ref{uone}) is the polynomial form 
of the Jacobi triple product
identity (see  page 49 of ref.\cite{and}).
When $L\rightarrow \infty$ 
the case $r=1,s=2$  
was first proven by Ramanujan in the 
famous ``lost notebook''(see eqn. 2.3.2 of ref.\cite{andb}) 
and the general case for the limits
$L\rightarrow \infty$ is a
consequence of the partition counting theorems 
of Andrews\cite{andc} and of Kadell\cite{kad}
on representations of $1/(q)_{\infty}$ by means of $r\times s$ Durfee
rectangles. The proof of the most general case with $L$ finite is obtained
by adapting the $L\rightarrow\infty$ proof to include  an upper bound on the
number and size of the parts of the partitions. Indeed a more general
result than (\ref{uone}) can be proven which generalizes 2.3.1 of
ref.\cite{andb}. The details of these
proofs will be published elsewhere.
The special case of $L\rightarrow \infty$ with all values
of $l$ summed together and $y=1$ was conjectured in ref. \cite{es}. 

To compare the fermi representations of the ${\hat u}(1)$ characters 
(\ref{uchar}) given by (\ref{uone}) and the representation of
ref. \cite{kmm} with the {\bf B} matrix (\ref{ubmat}) we first note that in
(\ref{uchar}) the case $p=p'=1~~R^2=2$ is the self dual point there
the characters are the same as the two characters of ${\widehat su}(2)_1.$
We note that the same ${\widehat su}(2)_1$ characters are obtained
from (\ref{uone}) with $r=1,~s=2$ in the $L \rightarrow \infty$ 
limit since $j\equiv L ({\rm mod}s).$ More generally we relate the two
fermionic forms by using $R^2=2p'/p=s/r.$ 

But the two different representations of $\chi_l^{(pp')}(q,y)$ given by
(\ref{ubmat}) and by (\ref{uone}) do not exhaust the representations
of the ${\hat u}(1)$ characters (\ref{uchar}) in terms of the universal
chiral partition function (\ref{ucpf}). For example the case $s=3,r=1$
has an $N=2$ supersymmetry \cite{gin} 
and for this case the three quasi particle
representation of the $c=1~N=2$ supersymmetric characters \cite{bmsa}
give a three quasi particle representation of the characters. 
For $R^2=2n$ and $n>3$ there is a representation
\cite{kkmma},\cite{kmm} with $n$
quasi particles and ${ \bf B}=2C^{-1}_{D_n}$ 
with $C_{D_n}$ the Cartan matrix of
the Lie algebra $D_n.$ The points $n=N^2$ with $N\geq 2$ are the
points shown in ref. \cite{gin} to have an extra ${\widehat su}(N)$ symmetry.

In
general for $R^2$ rational 
there is a
representation which originates 
in the Bethe's Ansatz solution to the
XXZ model of Takahashi and Suzuki \cite{ts}. This
representation is analogous to the representation of the characters of
the $M(p,p')$ minimal model proven in detail in
ref. \cite{bm},\cite{bmsb}. These
representations differ from the cases discussed above in that in
general most of  the parameters $\bf u$ in (\ref{ucpf}) are finite.

We thus see that there are at least three ways to describe the 
${\hat u}(1)$ compactified boson of rational radius in terms of exclusion
statistics. This multiplicity of representations at times leads to a
confusion of language, For example it is said in ref. \cite{iso} that
an anyon representation of the characters for $r^2=2$ was given in
ref. \cite{bls}. However the anyons of ref.\cite{iso} lead to the case
$r=1,s=2$ of (\ref{uone}) whereas the two representations of
ref. \cite{bls} are the representation (\ref{ubmat}) and a 
form with an infinite number of quasi particles first
proposed by Melzer \cite{mel}. It
could be argued that the term ``anyon'' should be reserved for the
excitations in (\ref{uone}) and the term ``spinon'' should be reserved
for (\ref{ubmat}). But there is no uniformity in the usage of these
terms and the term spinon is sometimes used to denote the
Takahashi-Suzuki representations as well. Indeed since for general
rational $R^2$ there are more that two representation
further new names are needed.

This multiplicity of fermionic representations (or equivalently the
multiplicity of possible exclusion statistics description)  is a very
important physical effect. In the case of the models $M(p,p+1)$
the different fermionic descriptions correspond to different massive
integrable perturbation (such as $\phi_{1,2},~\phi_{1.3}$ and $\phi_{2,1}.$)
But these different representations also correspond to different ways
of constructing the massless theory itself and these constructions may
be thought of as involving different ultraviolet regularizing
procedures needed to define the theory. In this interpretation we see
that the different regularizing procedures can lead to different
particle descriptions which are in general not local with respect to
one another. This phenomena is most important in the applications of
${\hat u}(1)$ to the fractional quantum Hall effect.

\section{Conformal field theory and Rogers-Ramanujan identities}

The fundamental principle behind the identification of the exclusion
statistics of Haldane with the fermionic counting rules
(\ref{count})--(\ref{mindef}) is the equivalence of the bosonic and
fermionic description of the underlying Hilbert space. This Bose/Fermi
equivalence is the principle behind all Rogers-Ramanujan identities and
is the reason why there is an equivalence between the fermionic
description of conformal field theory arising from the thermodynamic
Bethe's Ansatz and the corresponding bosonizations of Kac--Moody
algebras. In lattice statistical mechanics and in conformal field
theory this equivalence is well known\cite{kkmma}-\cite{suz}. 
However in the study of the
exclusion statistics on the fractional quantum Hall effect this
equivalence does not seem to be widely and explicitly recognized\cite{sch}. 
We will thus conclude with a few suggestions as to 
why this identification has not
previously been made.

There are perhaps two obvious obstacles to the identification of the 
the exclusion statistics of Haldane\cite {hald} with the fermionic
counting rules (\ref{count})--(\ref{mindef}) and the universal chiral
partition function (\ref{ucpf}).
The first is that in most applications of 
the universal chiral partition function (\ref{ucpf}) to conformal field theory 
only the special case $y_j=1$ occurs because in the CFT applications
there was no conservation law imposed on the number of excitations. 
In particular while Fermi/Bose (Rogers--Ramanujan identities) are known
for all minimal models $M(p,p')$ the bosonic form for the partition
function with a fugacity $y\neq 1$ is only known for the special cases
$M(2,2n+1)$ (see chapter 7 of the book of Andrews\cite{and}).

The second obstacle to identification 
is that in the definition of exclusion statistics of
Haldane\cite{hald} all values of $g_{\alpha \beta}$ are allowed whereas
in the applications of (\ref{ucpf}) to conformal field theory
only very specific values of the
matrices ${\bf B}$ are allowed. For example only three cases of the
scalar case $n=1$ are related to conformal field theories: $b=2$ is
$M(2,5),~b=1$ is $M(3,4)$ and $b=1/2$ is $M(3,5).$ We note that the
case $b=1/2$ is sometimes referred to as ``semionic'' and there is an
equality of the characters of $SU(2)_1$ and $U(1)_2$.  For these three
values of $b$ there are two special additional properties 
of (\ref{ucpf}) with $u=\infty$
and $y=1.$ First of all the dilogrithms in (\ref{usen}) are all
rational multiples of $L(1)$ and secondly (\ref{ucpf}) transforms
under a representation of the modular group. This second property is
of great importance in the theory of Kac--Moody algebras\cite{kac} and
conformal field theory\cite{ciz} 
In a similar fashion these three special cases seem to be the only
ones which are related to Bethe's Ansatz models\cite{abf}.
From this point of view we are here
proposing that the word ``universal'' in the universal chiral
partition function is to be used in a much wider sense than the
conformal field theory context where it first appeared.

\section{Conclusion}

We conclude with the statement that all of the identifications made
here of exclusion statistics of ref.\cite{hald} with 
the universal chiral partition
function (\ref{ucpf}) and the fermionic counting rules
(\ref{count})-(\ref{mindef})
 in section 2 and 3 for the scalar case can be extended
to the matrix case as well. Thus, since the fermionic 
counting rules (\ref{count})--(\ref{mindef}) are more general 
than the exclusion
statistics (\ref{haldef})  and they include  (\ref{haldef}) 
as the special case ${\bf u}=\infty,$ we propose that in one dimension
(\ref{count})-(\ref{mindef})
is a more natural and general definition of ``exclusion
statistics'' and in the future should be taken as the definition of the
term instead of (\ref{haldef}).

\acknowledgments 

We are grateful to acknowledge most useful discussions with
G. Andrews, R. van Elburg, P. Fendley, J. Jain, 
K. Schoutens, Y.S. Wu and D. Zagier. One of us (BMM)
wishes to thank G. Mussardo, A. Ludwig, A. LeClaire, and Y. Lu for
hospitality extended at the ICTP in Trieste where some of 
this work originated. 
This work is supported in part by the National Science Foundation
under grant DMR 97-03543.

\end{document}